# Study on neutron radiation field of carbon ions therapy

XU Jun-Kui(徐俊奎)[1,2], SU You-Wu(苏有武)[2], LI Wu-Yuan(李武元)[2], YAN Wei-Wei (严维伟)[2], CHEN Xi-Meng(陈熙萌)[1], MAO Wang[2](毛旺), PANG Cheng-Guo(庞成果)[1]

[1]School of Nuclear Science and Technology, Lanzhou University, Lanzhou 73000, China

[2]Institute of Modern Physics, Chinese Academy of Science, Lanzhou 730000, China

**Abstract** Carbon ions offer significant advantages for deep-seated local tumors therapy due to their physical and biological properties. Secondary particles, especially neutrons caused by heavy ion reactions should be carefully considered in treatment process and radiation protection. For radiation protection purposes, the FLUKA Code was used in order to evaluate the radiation field at deep tumor therapy room of HIRFL in this paper. The neutron energy spectra, neutron dose and energy deposition of carbon ion and neutron in tissue-like media was studied for bombardment of solid water target by 430MeV/u C ions. It is found that the calculated neutron dose have a good agreement with the experimental date, and the secondary neutron dose may not exceed 1‰ of the carbon ions dose at Bragg peak area in tissue-like media.

Key Word：Radiation protection, carbon ions therapy, energy deposition

## 1. Introduction

Heavy ions have been wildly used in various fields of nuclear physics, medicine, cancer therapy and so on. Especially heavy ions cancer therapy is more rapidly in recent years, this own to the physical characteristics of heavy ions beam. Heavy ions has a fixed range in the target matter, and characterized by a small entrance dose and a distinct maximum (Bragg peak) near the end of range. The position of this peak can be adjusted to the desired depth in tissue by changing the kinetic energy of the incident ions [1]. In the process of carbon ions cancer therapy, various kinds of secondary particles are created through nuclear spallation with the accelerator component and the body of patient. Secondary particles caused by carbon ions reactions with target must be carefully considered in treatment. Neutrons are the most abundant product in all secondary particles. Neutron cannot cause ionization directly, it affect a large area due to its strong penetrating power, so it may influence the surrounding environment and the whole body of the patient including both tumor and health tissue. Thus neutron radiation field is very important in safety evaluation in heavy ion therapy, and it is the fundamental measurable quantity required for all radiation protection in accelerators. This is because neutrons constitute the major component of prompt radiation produced during beam loss, and this radiation needs to be shielded for personnel, equipment and environmental protection against radiation hazard in such accelerator facilities. It is very important to evaluate the secondary neutron energy spectra, for the neutron dose estimation is according to the appropriate flux and dose conversion coefficients given by International Commission of Radiological Protection (ICRP) [2]. So, this fundamental study are important of estimating the probability of effect of the secondary particles produced during heavy ion therapy [3, 4].

Some researchers have been on neutron yield, spectra and angular distributions from heavy-ion bombarding of thick targets studying. The high neutron energy spectra and angular distribution of carbon ions bombard on thick copper target was studied by T. Nakamura at HIMAC (Heavy Ion Medical Accelerator in Chiba) in 1999[5]. More study of neutron yields of heavy ions hit thick target were studied [6-7]. In our previous work, Gui Sheng LI measured neutrons fluence from 50MeV/u oxygen ions on thick targets using the activation method at HIRFL of IMIP in 1999 [8]. The neutron dose distribution of 100MeV/u carbon ions therapy was measured in superficial tumors treatment terminal at HIRFL in 2010[9]. With the continuous development of heavy ion radiotherapy applications, now two HIMM (Heavy Ion Medical Machine) are building in Lanzhou and Wuwei Gansu respectively, which is described in detail in the previous work [10]. The neutron energy spectra, angular distribution, dose equivalent, fluence rate, etc. are very important for neutron physics, health physics, radiation protection studies and neutron shielding calculation.

In this work, the secondary neutron energy spectra was studied with carbon ions bombard on thick target with Monte Carlo program code, it is find that the secondary neutron energy spectra have a good agreement with previous experiment dates. At the same time, the distribution of neutron dose was calculated. The purpose of this work is to provide a full set of numerical data, which can be directly employed for shielding calculations of high-energy heavy ion accelerators and impacting of the patient within treatment process. Then the distribution of neutron flux in the tissue-like media and the surrounding space, the neutron contribution in heavy ion treatment process was studied, and we found that the contribution of neutron dose in the tissue-like media does not exceed 1‰ of the carbon dose in Bragg peck area.

2. Monte Carlo calculation

Various Monte Carlo transport codes have been utilized for the design of beam-transport lines and shielding at heavy ion accelerator facilities. For such transport program codes, it is required to simulate the nucleus-nucleus interaction in wide energy ranges. FLUKA [11-12] is the computer code which released as free software. FLUKA describes the nucleus-nucleus interaction using Dual Parton Model (DPM) [13] with energy over 5GeV/u, Relativistic Quantum Molecular Dynamics Model (RQMD) [14] with energy from 0.1GeV/u to 5GeV/u and Boltzmann Master Equation (BME) [15] theory below energy of 0.1GeV/u. The topological expansion, when supplemented with generally accepted theoretical principles like duality, unitarily, Reggae behavior and the parton structure of hadrons, provides the basis underlying the dual parton model (DPM). DPM provides a complete, phenomenological description of all facets of soft processes. This is a non-trivial achievement in view of the large amount of soft multiparticle data available from both hadronic as well as nuclear beams and targets. The RQMD is a complete transport theoretical scenario of nucleus-nucleus reactions, from the initial state of two nuclei before overlap to the final state after the strong inter actions have ceased (freeze-out). And RQMD is a semi classical microscopic approach which combines classical propagation with stochastic interactions. The Boltzmann Master equation (BME) theory is a nucleon transport model based on nucleon–nucleon collision processes in the nuclear potential and describes the statistical evolution of a sample of systems, which evolve from an initial state far from statistical equilibrium to an equilibrium state through a sequence of two body interactions and emission of unbound particles to the continuum.

Fig1 show the schematic diagram of HIFRL layout, the treatment room is the place where the neutron dose measured. The simulation geometry which used in this paper was only with shielding wall and target. Fig2 show the sectional drawing of calculation model. The beam cross section diameter is 2 mm. The copper target is a cylinder 5cm in diameter and 5 cm in thickness, and its axis coincides with beam-line. The water tissue-like target size is 30cm×30cm×30cm , and its component is listed in table1. The size of treatment room is 4m×6m×6m, filled with dry air.

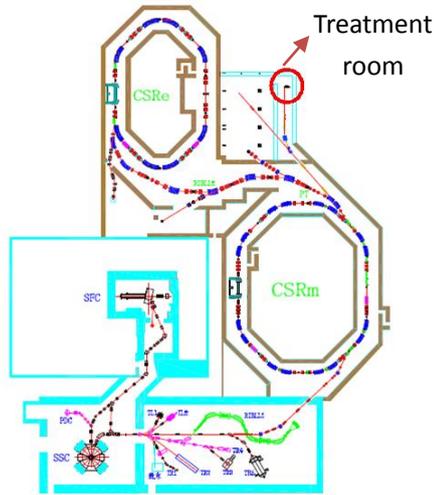

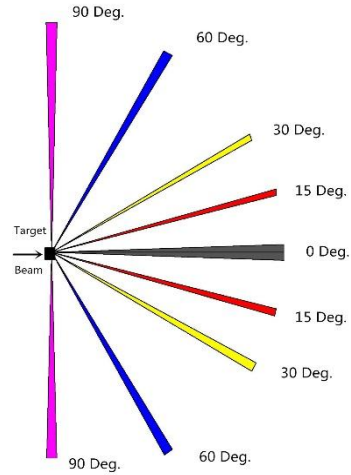

Fig.1 The schematic diagram of HIRFL

Fig.2. Spectra statistical method used in simulation

Table1 The component of tissue-like target

| component | H | C | N | O | Cl | Ca |
|---|---|---|---|---|---|---|
| Proportion (%) | 8.1 | 67.2 | 2.4 | 19.9 | 0.1 | 2.3 |

3. Result and discussion

In this work, if we using a geometry that is identical with the experimental setup, it is heard to obtain the good statistics due to the small detection area for particle counting. So we take small solid angle as the angle counting [16]. Fig.2 shows the geometry of the detectors modeled for calculation of neutron energy spectra, in every angle direction the solid angle used for counting is the same size. Before the second neutron energy spectra of carbon ions bombard on water target was calculated we calculate spectra with copper target. Fig 3. Shows the neutron spectra from copper target bombarded with 400Mev/u carbon ions. The FLUKA result agree well with the experimental date which measured by T. Nakamura [5], because of the error for initial ions counting we use the relative results in the plot. Therefor the results that we got is believed reasonable.

The neutron energy spectra was calculated from solid water target bombarded with 430MeV/u carbon ions is shown in Fig.4, and the neutrons dose equivalent distribution was presented in Fig.5, and Fig.6 present the angle distribution of neutron does comparison between the FLUKA result and our experimental date measured at deep tumor therapy terminal of IMP, the distance from detector to target is 2 meter, the instruments used in the experiment is Wendi-II neutron dose-meter with model of 579.

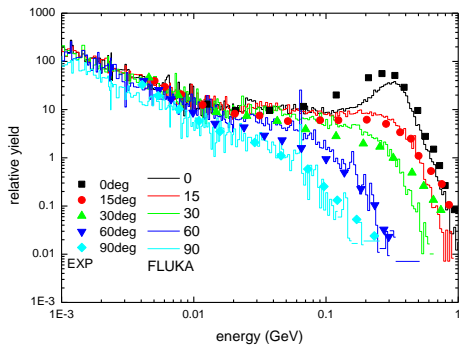

Fig.3 Comparison between the result and the experiment date for 400MeV/u carbon ions bombardment of copper target

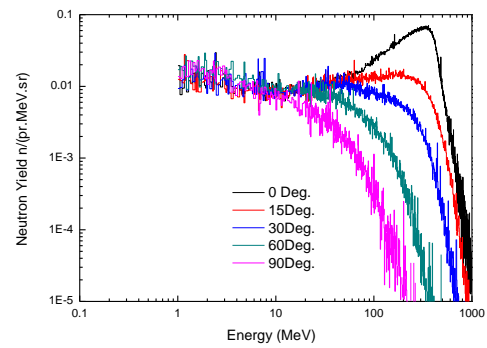

Fig.4 The neutron spectra is simulated with FLUKA for 430MeV/u carbon ions bombardment of solid water target

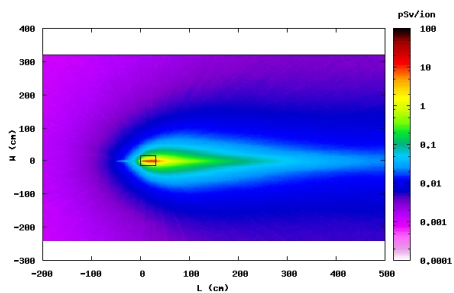

Fig.5 The spatial distribution of neutron dose

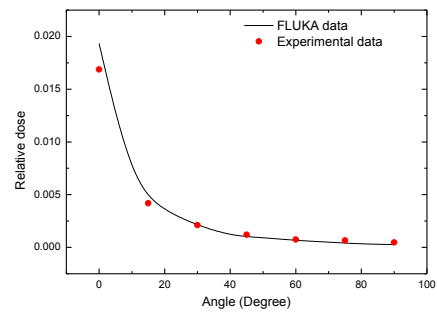

Fig .6 The angular distribution of neutron dose

Fig3 and Fig.4 indicate that there is a broad peak at high energy in the neutron spectra, especially a large peak at 0 degree. The neutrons energy in the forward direction achieve up to about the twice of the incident ions energy per nucleon. Fig.3 shows that the peak energy is about 67% of the beam ion per nucleon, 270MeV for 400MeV/u carbon ions bombard on copper target. Fig.4 shows that the peak energy is about 81% of the projectile energy per nucleon, 350MeV for 430MeV/u carbon ions bombard on solid water target. The neutron spectra become harder for light target nuclei, this because the sum of the transferred energies is greater for lighter target nuclei. Those high energy neutrons component produced in the forward direction by spallation process which is called as cascade process. Low neutrons are mainly produced in compound nucleus system by evaporation process, so the neutron emission by spallation process has a sharp peaking of angular distribution. In large angel, the neutron spectra become softer because the evaporation process is dominated. Fig 5 shows the spatial distribution of neutron dose with unit of pSv/ion, Fig6 shows the angle distribution of neutron dose, the solid point is the experiment value and the line is FLUKA date, due to calibration error of the detector for counting primary ions, therefore, relative dose was used in the figure. Fig.5 and Fig.6 has a similar characteristic with the neutron spectra, the neutron dose is the largest at 0 degree, the dose decreases with the angle increase. This can be explained that the energy spectra can be converted to ambient and personal dose equivalents by folding with the appropriate flounce to dose conversion coefficients given by International Commission of Radiological Protection.

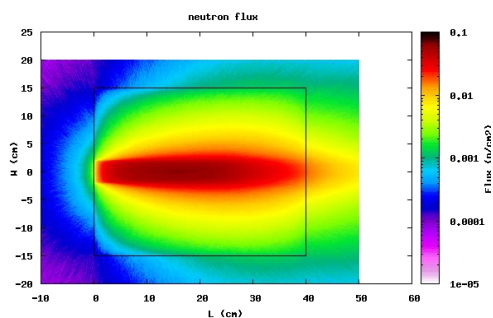 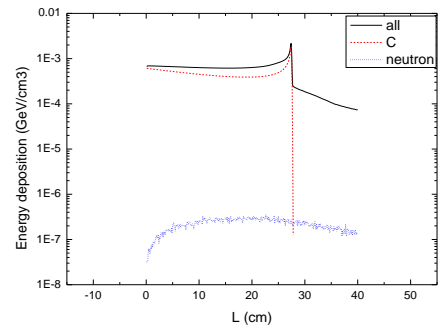

Fig.7. Neutron flux in and around the solid water target

Fig.8 Energy deposition for 400MeV/u carbon ions bombard on solid water target

Fig.7 shows the neutron flux density distribution in and out the water target. Along the beam direction in the target, neutron yield increased first and then decreased, and neutron yield is lower in the place where lateral from the beam direction. Fig.8 shows that energy deposition when 400MeV/u carbon ions bombarded on solid water target with unit in GeV/cm$^3$, the solid line represent the all energy deposition, dash line shows the carbon energy deposition, and the dot line is the neutron energy deposition. The potion between solid and dash line, especially beyond the Bragg peak are entirely due to the contribution of the fragmentation which come from nuclear reactions. Energy deposition from secondary neutrons is analyzed as following. Neutrons as neutral particles do not lose their energy directly via ionization, but rather via the secondary reactions involving charged hadrons, they can be produced either in neutron elastic scattering on target nuclei or via inelastic nuclear reactions. However, there is few publication reporting the doses from secondary neutrons inside the extended target or phantom in heavy ion irradiations. Since a measured depth-dose distribution gives the total dose, it includes the local energy deposition from charged hadrons and nuclear fragments from neutron induced reactions. So the Monte Carlo simulation calculation is the best way to study second neutron energy deposition. In our simulation result (Fig.8), the neutron energy deposition may not exceed 1‰ of the carbon ions dose at Bragg peak area in tissue-like media.

4. Conclusion

In this paper, the secondary neutron spectra with copper target was calculated with FLUKA of version 2011.2c, the result was in overall agreement with the previous experimental dates. Then the secondary neutron energy spectra and dose distribution with the treatment environment was calculated. And the angle distribution of neutron dose is well agreement with the experimental values which measured at deep tumor treatment room of IMP by us. These data are very important for accelerator shielding and individual dose assessment. These studies show that secondary neutron energy deposition should be smaller than the total energy deposition. It is demonstrated that the additional neutrons produced by carbon ion beam do not appear to be a serious hazard to human body in the therapeutic process.